\documentclass[journal]{IEEEtran}
\usepackage{latexsym,,amssymb,amsmath,graphicx,epsf,cite,bbm,float}
\usepackage{ifpdf}
\usepackage{epstopdf,mathtools}

\usepackage{algorithm,algorithmic}
\usepackage{amsmath,amssymb,bm}
\usepackage{amsfonts,dsfont,color,bbm,subcaption}

\def\ninept{\def\baselinestretch{1}}
\ninept

\DeclareMathOperator*{\argmax}{arg\,max}
\DeclareMathOperator*{\argmin}{arg\,min}

\usepackage{hyperref,amsthm}

\newtheorem{remark}[]{Remark}

\newtheorem{example}[]{Example}

\begin{document}

\title{Robust, Nonparametric, Efficient Decomposition of Spectral Peaks under Distortion and Interference} 
\author{\IEEEauthorblockN{Kaan Gokcesu}, \IEEEauthorblockN{Hakan Gokcesu} }
\maketitle

\begin{abstract}
	We propose a decomposition method for the spectral peaks in an observed frequency spectrum, which is efficiently acquired by utilizing the Fast Fourier Transform. In contrast to the traditional methods of waveform fitting on the spectrum, we optimize the problem from a more robust perspective. We model the peaks in spectrum as pseudo-symmetric functions, where the only constraint is a nonincreasing behavior around a central frequency when the distance increases. Our approach is more robust against arbitrary distortion, interference and noise on the spectrum that may be caused by an observation system. The time complexity of our method is linear, i.e., $O(N)$ per extracted spectral peak. Moreover, the decomposed spectral peaks show a pseudo-orthogonal behavior, where they conform to a power preserving equality.
\end{abstract}

\section{Introduction}

The parameter estimation for a mixture of sources (i.e., the amplitudes, frequencies and phases for a mixture of sinusoids) is a prominent problem in the fields of signal processing \cite{luo2015frequency}; especially in power systems \cite{lovisolo2005efficient,grainger1999power,akers2006hydraulic}, communication systems \cite{schwartz1995communication,stremler1990introduction,tse2005fundamentals}, acoustic analysis \cite{titze1994toward,fischer1954acoustic,drioli2003emotions} and anomaly detection \cite{chandola2009anomaly,gokcesu2017online,patcha2007overview,delibalta2016online,ahmed2016survey,gokcesu2018sequential}. This problem is encountered in a myriad of real life applications such as radar imaging \cite{wang2016parameters,selva2005efficient}, wireless communications \cite{luo2016interpolated,luo2015frequency}, spectroscopy analysis \cite{duda2011dft,umesh1996estimation}, vibration analysis \cite{norton1990fundamentals,petyt2010introduction,brandt2011noise} and biosignal analysis \cite{akay2012biomedical,chang2010biomedical}. The biggest issue that arises in these real world applications is the deformation of the original signal by noise, interference or any distortion present in the application, which creates a need for robust methods \cite{steinhardt2018robust,gokcesu2021generalized,gokcesu2022nonconvex}.

Especially in the field of sequential analysis (i.e., time-series) \cite{gokcesu2021nonparametric,hamilton2020time}, the parameter estimation of a mixture of sinusoids have been heavily studied \cite{stoica1997introduction,zielinski2011frequency}. In general, these algorithmic approaches of spectral peak analysis can be broken down into two main categories: parametric and non-parametric.

The traditional nonparametric approaches include many distinct algorithms \cite{capon1969high,li1996adaptive,yardibi2010source}, where the goal is the accurate estimation of the signal spectrum. The algorithms extract the spectral peaks from the spectrum estimate by straightforward maximization. Although the spectrum estimation of these algorithms are highly accurate, their computational cost is exhaustive. The time complexity is quadratic in the number of samples $N$ in the most efficient implementation, i.e., $O(N^2)$, and may even be much higher \cite{glentis2008fast,xue2011iaa,angelopoulos2012computationally}.

Parametric estimation \cite{gokcesu2017density} techniques are also widely considered in the literature. In contrast to the nonparametric ones, instead of estimating the spectrum, the approaches utilize the limited modeling of the sinusoids and estimate their specific parameters. These methods can be categorized under two schools of thought, which are frequency and time domain algorithms \cite{belega2010accuracy}.

The time domain algorithms are very popular because of their strength in modeling the sinusoidal parameters. They generally utilize matrix based methods to create powerful algorithms such as \cite{hua1990matrix,vanhuffel1994algorithm,haardt1995unitary,schmidt1986multiple,pisarenko1973retrieval}. The matrix based approaches aim to split the observed signal domain into source and noise subspaces, generally by using SVD based techniques \cite{wall2003singular}. Because they utilize matrix decomposition \cite{abdi2007eigen} and inversion, their complexity is comparatively high. 
To increase efficiency, iterative methods are also used such as \cite{bresler1986exact,chan2010efficient,sun2012efficient}, where the goal is the minimization of some error between the estimation and observation under varying constraints. Nonetheless, all of these time domain methods suffer from the high computational costs one way or the other because of the matrix operations, which are at least quadratic in the number of samples \cite{boyd1994linear}. This high computational cost results from the relatively harder optimization of the frequencies, which, in general, is a nonconvex problem that requires global optimization techniques \cite{henrion2004solving,gokcesu2021regret,gokcesu2022low}. 
While some methods circumvent this problem by utilizing the specific structure of the operated matrices \cite{selva2005efficient, klinger1967vandermonde}, their computational complexity can be exceptionally high depending on the observation and its source model.

All in all, because of the computational complexity concerns, the frequency domain approaches are much more popular. This is a simple result of the fact that the Discrete Fourier Transform of samples can be calculated efficiently in linearithmic time with Fast Fourier Transform \cite{nussbaumer1981fast}. There exist many methods utilizing this efficiency such as \cite{gough1994fast,li1996adaptive,gokcesu2022blind}, which work in an iterative alternating procedure, where the model parameters are estimated and subtracted from the residual signal. They generally work by maximizing the periodogram of the observed signal which can result in high estimation error \cite{aboutanios2011estimating}. To address this issue, many methods have been proposed \cite{quinn1994estimating,aboutanios2005iterative,yang2011noniterative,selva2011efficient}.
Nonetheless, they are specifically designed for the estimation of a single sinusoid and their applicability in a mixture is questionable because of interference between different signal sources. This interference is a direct result of the spectral leakage and we can only observe the spectrum on the DFT values, which creates the picket fence effect \cite{santamaria1998improved,zhang2001algorithm}. To this end, joint optimization techniques have also been proposed \cite{ye2017rapid,gokcesu2022blind}. Moreover, the interference reducing methods are also a rich field of study \cite{diao2013interpolation,duda2014interpolated,duda2011dft,luo2015frequency,luo2016interpolated}, where the general approach is to apply a windowing function \cite{gokcesu2022smoothing,prabhu2014window} to limit the leakage. However, such windowing functions can result in decreased accuracy in frequency estimation because of the picket fence effect \cite{ye2017rapid}.

In this work, we combine the accuracy of the nonparametric techniques and efficiency of the frequency domain approaches to create a robust decomposition algorithm.

\section{Preliminaries}\label{sec:prelim}
We have the real observations $\boldsymbol{y}=\{y_n\}_{n=0}^{N-1}$. Let these observations be the output of a system function such that
\begin{align}
	y_n=\mathcal{H}(x_n),
\end{align}
where $x_n$ is the source and $\mathcal{H}(\cdot)$ collectively represents the noise on the input including interference and distortion. Let $x_n$ be a mixture of sinusoids (i.e., originating from a multi-tone signal) such that
\begin{align}
	x_n=\sum_{r=1}^{R}A_r\cos(w_rn+\theta_r),
\end{align}
for some $R$ number of sinusoid components; where the amplitude, frequency and phase of the $r^{th}$ component is given by $A_r$, $w_r$ and $\theta_r$ respectively.

There exist many techniques to extract the parameters of these sinusoids given the input values $x_n$, especially when the number of components $R$ is known a priori.

Specifically, the optimization of the amplitude $A_r$ and phase $\theta_r$ is relatively easier because of the linear dependency on the estimation. The main issue, in general, is the estimations of the frequencies. For this problem setting, most prominent approaches generally utilize the power spectrum; and the approach boils down to estimating the frequency as the most dominant values in the power spectrum.

Although there exists many approaches for the estimation of the power spectrum, the techniques involving Fast Fourier Transform are more popular because of their sheer efficiency.
The DFT of the input $x_n$ is given by
\begin{align}
	X_k=\sum_{n=0}^{N-1}x_ne^{-j\frac{2\pi kn}{N}},
\end{align}
where $k\in\{0,\ldots,K\}$ and $K=\lfloor N/2\rfloor$ (single side of DFT). With efficient FFT methods, $\{X_k\}_{k=0}^K$ can be calculated in linearithmic time, i.e., $O(N\log N)$. To detect sinusoids in this spectrum, we simply need to locate the spectral peaks. Originally, the sinusoidal frequencies should have a delta function effect on the spectrum. However, because of the limited number of samples (multiplication by rect window), the spectrum is convolved by a sinc function and we need to fit it accordingly (spectral leakage and picket fence effect).

However, since we do not have access to the input $x_n$, we need to analyze the FFT of $y_n$, which is given by
\begin{align}
	Y_k=\sum_{n=1}^{N}y_ne^{-j\frac{2\pi kn}{N}},
\end{align}
where $k\in\{0,\ldots,K\}$ and $K=\lfloor N/2\rfloor$ (single side of FFT). Because of the system function $\mathcal{H}$, the problem becomes much harder. Although Additive White Gaussian Noise (AWGN) can be somewhat tolerated, the system function $\mathcal{H}$ may distort the signal beyond measure. Some real world examples include the windowing of the input and truncation of the amplitudes. To this end, simply fitting a sinc or other windowing artifacts may not be robust enough and estimations may be error prone. Hence, we fit a more general class of functions.

\section{Robust Decomposition of Spectral Peaks}

In this section, we detail our methodology. We start by proposing the notion of pseudo-symmetric waveform as a model for the spectral peaks.

\subsection{Pseudo-Symmetric Spectral Peaks}

The original approach in finding the spectral peaks is analyzing the peak values in the FFT $Y_k$ of $y_n$. However, for there to be no interference, the frequency should be an integer multiple of $2\pi/N$, which is generally not the case. Moreover, we would have observed a delta function located at the frequency if our input length was infinite. However, because of the limited number of samples, we only observe the infinite length signal multiplied by a rectangle window. Thus, the original spectrum is convolved with the window FFT. When $N$ is large enough this convolution is approximately a sinc function, i.e.,
\begin{align}
	S(k)=\frac{\sin(\pi k)}{\pi k},
\end{align}
for $k\in\Re$; which is symmetric around $k=0$, $S(0)=1$ and $S(k)=0$ when $k$ is integer. When the original spectrum is convolved with it, if the central frequency coincides with an integer multiple of $2\pi/N$, its effect on the other FFT values will be zero. Otherwise, we will have the spectral leakage. Since we are only dealing with FFT instead of DTFT, we will have the picket fence effect, where we only observe the contribution on the samples separated by integer multiples of $2\pi/N$.

Suppose we want to fit a sinc model centered around the frequency $k^*+\delta$, where $k^*\in\{0,\ldots,K\}$ and $\delta\in[-1/2,1/2]$. The contribution on FFT for any $k$ will be
\begin{align}
	S_{k^*+\delta}(k)=&\frac{\sin(\pi(k-k^*-\delta))}{\pi(k-k^*-\delta)},\\
	=&\frac{\gamma}{\pi|k-k^*-\delta|},
\end{align}  
where $\gamma=|\sin(\pi\delta)|$. However, this model is too restrictive and may be error prone because of the distortion and interference of the transfer function $\mathcal{H}$. To this end, we fit a more robust function instead of the sinc function. First of all, we observe that the original sinc function has an interesting behavior, which is
\begin{align}
	S_{k^*+\delta}(k^*)\geq S_{k^*+\delta}(k^*+1)\geq S_{k^*+\delta}(k^*-1)\geq\ldots\\
	S_{k^*+\delta}(k^*)\geq S_{k^*+\delta}(k^*-1)\geq S_{k^*+\delta}(k^*+1)\geq\ldots
\end{align}
depending or whether $\delta\geq 0$ or $\delta\leq 0$. To this end, for increased robustness, we do not assume any functional form on our model but instead assume that it is pseudo-symmetric, i.e., it has one of the following two forms:
\begin{align}
	P(k^*)\geq P(k^*+1)\geq P(k^*-1)\geq \ldots\\
	P(k^*)\geq P(k^*-1)\geq P(k^*+1)\geq \ldots
\end{align}
Fitting a symmetric model is not intuitive because of the misalignment and the picket fence effect.

\subsection{Fitting a Pseudo-Symmetric Peak}\label{sec:pseudo}

We have the FFT spectrum $\{Y_k\}_{k=0}^K$ to fit a pseudo-symmetric peak, a center bin $k^*\in\{0,\ldots,K\}$ and direction $b\in\{-1,1\}$. We want to construct a spectrum $\{Z_k\}_{k=0}^K$ such that it is pseudo-symmetric with $k^*$ and $b$, i.e.,
\begin{align}
	|Z_{k^*}|\geq|Z_{k^*+b}|\geq|Z_{k^*-b}|\geq|Z_{k^*+2b}|\geq\ldots
\end{align}
Hence, we are constructing the monotonicity in an alternating manner around $k^*$ by starting with the index $k^*$ and $k^*+b$. When we reach a boundary on the spectrum, i.e., $k^*\pm ib\in\{0,K\}$ for some $i$ we continue on from the otherside and end the sequence when we reach both boundaries.

\begin{example}
	Given $K=10$, $k^*=3$, $b=1$; we have the following index sequence
	\begin{align*}
		\mathcal{I}=\{3,4,2,5,1,6,0,7,8,9,10\}.
	\end{align*}
\end{example} 

After the creation of the index sequence, we have the fitted variables $Z_k$ and the target variables $Y_k$. By rearranging with the index sequence, we have a problem of isotonic regression \cite{gokcesu2021efficient,gokcesu2021optimally}.

The specific optimization problem is given by 
\begin{align}
	\argmin_{\{Z_k\}_{k=0}^K}&\sum_{k=0}^{K}l(Z_k,Y_k)\\
	&|Z_{\mathcal{I}(m)}|\geq|Z_{\mathcal{I}(m+1)}|, m\in\{0,\ldots,K-1\},
\end{align} 
for some index sequence $\mathcal{I}$ and loss function $l(\cdot,\cdot)$.
We use the squared error as our loss function, i.e., $l(Z_k,Y_k)=|Z_k-Y_k|^2$. Hence, the problem becomes
\begin{align}
	\argmin_{\{Z_k\}_{k=0}^K}&\sum_{k=0}^{K}|Z_k-Y_k|^2,\\
	&|Z_{\mathcal{I}(m)}|\geq|Z_{\mathcal{I}(m+1)}|, m\in\{0,\ldots,K-1\}.
\end{align} 
Since the nonincreasing constraints are on the magnitudes of $Z_k$, we can further revise the problem as
\begin{align}
	\argmin_{\{|Z_k|,\theta_k\}_{k=0}^K}&\sum_{k=0}^{K}||Z_k|e^{j\theta_k}-Y_k|^2,\\
	&|Z_{\mathcal{I}(m)}|\geq|Z_{\mathcal{I}(m+1)}|, m\in\{0,\ldots,K-1\}.
\end{align} 
For least squares optimization, given the magnitudes, the minimizer phases are given by the phases of the target variables. Hence, the problem becomes
\begin{align}
	\argmin_{\{|Z_k|\}_{k=0}^K}&\sum_{k=0}^{K}||Z_k|-|Y_k||^2,\\
	&|Z_{\mathcal{I}(m)}|\geq|Z_{\mathcal{I}(m+1)}|, m\in\{0,\ldots,K-1\}.
\end{align} 
This problem can be solved in $O(K)$ time \cite{gokcesu2021efficient} and provides us with the magnitudes $|Z_k|$ and $Z_k$ can be constructed straightforwardly as 
\begin{align}
	Z_k=\begin{cases}
		\frac{Y_k}{|Y_k|}|Z_k|,& |Y_k|>0\\
		|Z_k|,& |Y_k|=0
	\end{cases}
\end{align}

\subsection{Central Frequency Bin}\label{sec:central}

To fit a pseudo-symmetric peak, we first need to extract the central frequency $k^*$ and the direction $b$.

One approach is the minimization of the loss with respect to $k^*$ and $b$. Let $\mathcal{I}_{k^*,b}$ be the index ordering implied by the central bin $k^*$ and the direction $b$. Then, the optimization problem becomes 
\begin{align}
	&\argmin_{k^*,b,\{|Z_k|\}_{k=0}^K}\sum_{k=0}^{K}||Z_k|-|Y_k||^2,\\
	&|Z_{\mathcal{I}_{k^*,b}(m)}|\geq|Z_{\mathcal{I}_{k^*,b}(m+1)}|, m\in\{0,\ldots,K-1\}
\end{align} 
However, this approach will have quadratic time complexity, where there exists $2K$ candidates. Therefore, efficient methods to determine $k^*$ and $b$ are needed. There are a number of things that can be considered in accordance with the problem setting and needs at hand.

First of all, if the goal is the detection of spectral peaks, however narrowband they are, we can directly use the maximum absolute valued FFT value, i.e.,
\begin{align}
	k^*=\argmax |Y_{k}|.
\end{align}
After determining $k^*$, since there exist at most $2$ candidates for $b$, we can solve for both of them and choose the fit that results in the lesser error.
Note that, this approach of determining $k^*$ is equivalent to finding the $M-1$ consecutive FFT samples with most power in a $M$ sample FFT recursively starting with $M=K$ FFT values. Similarly, any window of $M-l$ (where $l$ is not necessarily $1$) can also be utilized. 

Moreover, instead of the linear elimination of the frequency bins, we can consider a multiplicative elimination recursively. For example, we can search for $M/2$ most consecutive power in $M$ sample FFT. In a sense, we recursively search for the most power containing half-band. This search can be done by calculating the cumulative power in the first $M/2$ frequency bins and iteratively updating it via sliding the window by $1$. In this way, the most power containing half-band can be found in $O(M)$ time. Since we start from $M=K$ and iteratively halve the frequency band, total complexity will be $O(K)$. 

Furthermore, instead of the most power containing half-band, we can search for the smallest band containing at least half the power. To do this, we can sum the power density starting from the first sample until the first time we exceed half the power. This will provide us with the bandwidth of the half-power spectrum starting at $0$. Then, we iteratively subtract the power densities from the beginning until we fall short of half the power. After that, we continue adding the power density of the samples until we exceed half power again. This approach will iteratively create a list of bandwidth that starts with each FFT index. Then, we can search this list to find the smallest band (with appropriate tie breakers), which will take $O(M)$ complexity in total. The bandwidth of this smallest band region is upper bounded by the bandwidth that is starting at $0$ and bandwidth that ends in $M$. Hence, the length is upper bounded by $(M+1)/2$, which makes the total time complexity $O(K)$.

\subsection{Iterative Decomposition}

To iteratively decompose the spectral peaks in the observed FFT $\{Y_k\}_{k=0}^K$, we propose the following procedure:
\begin{enumerate}
	\item Set input to $W_k=Y_k,\forall k$ 
	\item Set $r=1$
	\item Detect the central frequency $k^*$ with inputs $W_k$ according to \autoref{sec:central}\label{step:ite}
	\item Fit a pseudo-symmetric spectral peak $Z_{k,+}$ with inputs $k^*$ and $b=1$ according to \autoref{sec:pseudo}
	\item Fit a pseudo-symmetric spectral peak $Z_{k,-}$ with inputs $k^*$ and $b=-1$ according to \autoref{sec:pseudo}
	\item Let $E_+=\sum_{k=0}^{K}|Z_{k,+}-W_k|^2$ 
	\item Let
	$E_{-}=\sum_{k=0}^{K}|Z_{k,-}-W_k|^2$
	\item If $E_+\leq E_{-}$, set $Z_k^{(r)}=Z_{k,+}$\\
	else, set $Z_k^{(r)}=Z_{k,-}$\label{step:b}
	\item Set $W_k\leftarrow W_k-Z_k^{(r)}$, $\forall k$
	\item Set $r\leftarrow r+1$
	\item If $\mathcal{C}(\{W_k\}_{k=0}^K)$ is TRUE, stop\\
	else, return Step \ref{step:ite}\label{step:end}
\end{enumerate}

\begin{remark}
	The function $\mathcal{C}(\cdot)$ at Step \ref{step:end} is the stopping criterion; which can be set in accordance with the application settings. If always kept false, the algorithm runs as an anytime algorithm.
\end{remark}

The advantages of using square loss when fitting the pseudo-symmetric spectrum in \autoref{sec:pseudo} is numerous from its computational efficiency to real world similarities. Moreover, we observe that when square loss is used in fitting, we acquire pseudo-orthogonal signals.

When fitting $Z_k$ to $W_k$, phase is kept the same and only magnitudes are optimized. In this isotonic regression problem, some set of $Z_k$ values are jointly set as the mean of some set of $W_k$. Hence, the residual is orthogonal, i.e., the final $Z_k$ and $W_k-Z_k$ are orthogonal to each other. Therefore, We have a power preservation equality, where
\begin{align}
	\sum_{k=0}^K |W_k|^2=\sum_{k=0}^K |Z_k|^2+\sum_{k=0}^K|W_k-Z_k|^2.
\end{align}
Since this equality perseveres in each iteration of the algorithm, we have the following at any arbitrary iteration $R$
\begin{align}
	\sum_{r=1}^{R}\sum_{k=0}^K|Z_k^{(r)}|^2+\sum_{k=0}^K|Y_k-\sum_{r=1}^{R}Z_k^{(r)}|^2=\sum_{k=0}^{K}|Y_k|^2,
\end{align}
where $Z_k^{(r)}$ is the fit at iteration $r$. Therefore, it is intuitive to select a stopping criterion based on the residual power $\sum_{k=0}^K|Y_k-\sum_{r=1}^{M}Z_k^{(r)}|^2$. Hence, we have
\begin{align}
	\mathcal{C}(\{W_k\}_{k=0}^K)=\begin{cases}
		TRUE,&\sum_{k=0}^K|W_k|^2\leq P_{\tau}\\
		FALSE,&\sum_{k=0}^K|W_k|^2> P_{\tau}
	\end{cases},
\end{align}
where $P_\tau$ is an appropriate power threshold. For a nonparametric design, this threshold can be chosen in accordance with the power of the original observed spectra, e.g., its mean. Hence,
\begin{align}
	P_\tau=\frac{1}{K+1}\sum_{k=0}^{K}|Y_k|^2.
\end{align} 

\section{Conclusion}\label{sec:disc}
We proposed a decomposition method for the spectral peaks in an observed frequency spectrum. Our method is efficient and utilizes the widely used FFT approach. In contrast to the traditional methods of fitting specific waveforms, we optimize the problem from a more robust perspective where we model the peaks in spectrum as pseudo-symmetric functions, where the only constrained is the nonincreasing behavior around a central frequency when the distance increases. Our approach is more robust against arbitrary distortion and interference on the spectrum that may be caused by an observation system function. It is efficient, since the time complexity is linear, i.e., $O(N)$ per extracted spectral peak. Given that we extract $R$ peaks, the total complexity is $O(RN)$. Moreover, the decomposed spectral peaks have a power preserving equality, i.e., pseudo-orthogonal.

\bibliographystyle{IEEEtran}
\bibliography{double_bib}
\end{document}